**Title**

ULTRA-HIGH NUMERICAL APERTURE WAVEGUIDE-INTEGRATED META BEAM SHAPER

**Authors**

Hrishikesh Iyer and Yurii Vlasov*

**Affiliations**

University of Illinois Urbana Champaign

Department of Electrical and Computer Engineering.

**CONTACT INFO**

*Corresponding author: Yurii Vlasov. Email: yvlasov@illinois.edu

**This PDF file includes:**
    Main Text
    Figures 1-6
    Methods
    References
    Supplementary Materials

**ABSTRACT**

Integration of metasurfaces with guided mode sources like waveguides have opened new frontiers for on-chip optical integration. However, the state-of-the-art in the field has targeted applications where long focal distances over thousands of light wavelengths are needed. This regime where the paraxial approximation holds enables inverse design of the metasurfaces with weakly confining elements that are typically thicker than the wavelength in the material. For short-focal length applications at distances less than 100$\lambda$ where the paraxial approximation fails, and high numerical apertures (NA) are necessary, a different approach is required. Here we designed and experimentally demonstrated single-mode waveguide-integrated meta beam shapers capable of redirecting the confined light into the free space and focusing it at focal distances less than 100$\lambda$ above the chip surface into a tightly focused spot. Focal spot characteristics measured at 460nm operating wavelength are approaching diffraction limited focusing across a range of focal lengths, device footprints, and numerical apertures demonstrating the robustness of our approach. Focal volumes smaller than 1$\mu$m$^3$ are demonstrated for a range of focal distances below 50$\mu$m. For a device with some of the highest NA amongst integrated metasurfaces of 0.95 the measured focal volume is as small as just 0.06$\mu$m$^3$ at a focal distance of 13$\mu$m. These on-chip integrated ultra-high NA meta beam shapers have the potential to unlock new applications in quantum optical computing with trapped ions, localized optogenetic neurostimulation, and high resolution in-situ microscopy.

## INTRODUCTION

The concept of a metasurface composed of arrays of subwavelength elements (meta-atoms) to achieve superior control over phase, amplitude and polarization of the radiation transmitted at normal incidence have made a profound impact on approaches for arbitrary beam forming, wave-front shaping, and holography[1,2] (**Fig.1A**). Recently, the ideas of integrating the metasurface elements into a planar waveguiding optics[3] to feed them with the in-plane guided modes and to control out-of-plane wavefronts radiated into a free space have received growing attention[4-19] (**Fig.1B**). Waveguide-fed metasurfaces have been demonstrated to enable 1D[11,16] and 2D[4,10] focusing, beam collimation[5,17], vortex beam generation[6], and holographic projection[14,15,18,19]. While initial demonstrations relied mostly on separate metasurface and waveguide designs aligned on top of each other[5], more recent developments considered meta-atoms placed within reach of the evanescent field of the guided mode[7,10,11,14] to manipulate separately four degrees of freedom (DOF) of scattered waves including phase, amplitude, and polarization states. Further development of structural manipulation of individual meta-atoms[6,9,12] to impose artificial birefringence and to break the symmetry of the otherwise periodic metasurface array enabled simultaneous and independent manipulation of all four optical DOF. These developments rely heavily on the concept of quasi-bound states in the continuum[20,21] that implies the non-local nature of the propagating mode which couples quasi-bound states between individual meta-atoms to arbitrarily tailor the out-of-plane wave fronts. This versatile approach is promising for many important applications including optical imaging, holography, quantum technology, and laser ranging that all are aiming at distances beyond a few thousand wavelengths (**Fig.1C**). At these distances, the beam shaping is achieved at angles near-normal to the surface with effective numerical apertures (NA) below 0.37. The design of such non-local metasurfaces is therefore simplified using a paraxial approximation to decouple the real and imaginary components of the scattered wave and to achieve full control of the scattered wavefront[9,12]. Many applications, however, require highly converging beams including large-angle light collection in high resolution microscopy, optogenetics applications[22-26], photolithography, high-density data recording[27], and in optical quantum computing with trapped ions[28-30]. These applications would benefit from the development of waveguide integrated meta beam shapers with high NA approaching 0.9 to control free-space beams at distances below 100 wavelengths above the surface of the photonic integrated chip (**Fig.1D**). At these ultrahigh NA the paraxial approximation fails, and non-locality might not be advantageous[31] to achieve high focusing efficiency[32].

Here we show, that using a high refractive index single mode waveguide with meta-atoms etched directly into its core can effectively focus the in-plane guided mode into a near diffraction limited spot at ultra-short distances above the chip surface with ultrahigh NA over 0.9. High index contrast results in diminishing interaction among meta-atoms to achieve rapidly varying spatial phase profiles enabling mode conversion from high-confinement in-plane-guided mode into a wide-angle 144° converging free-space beam.

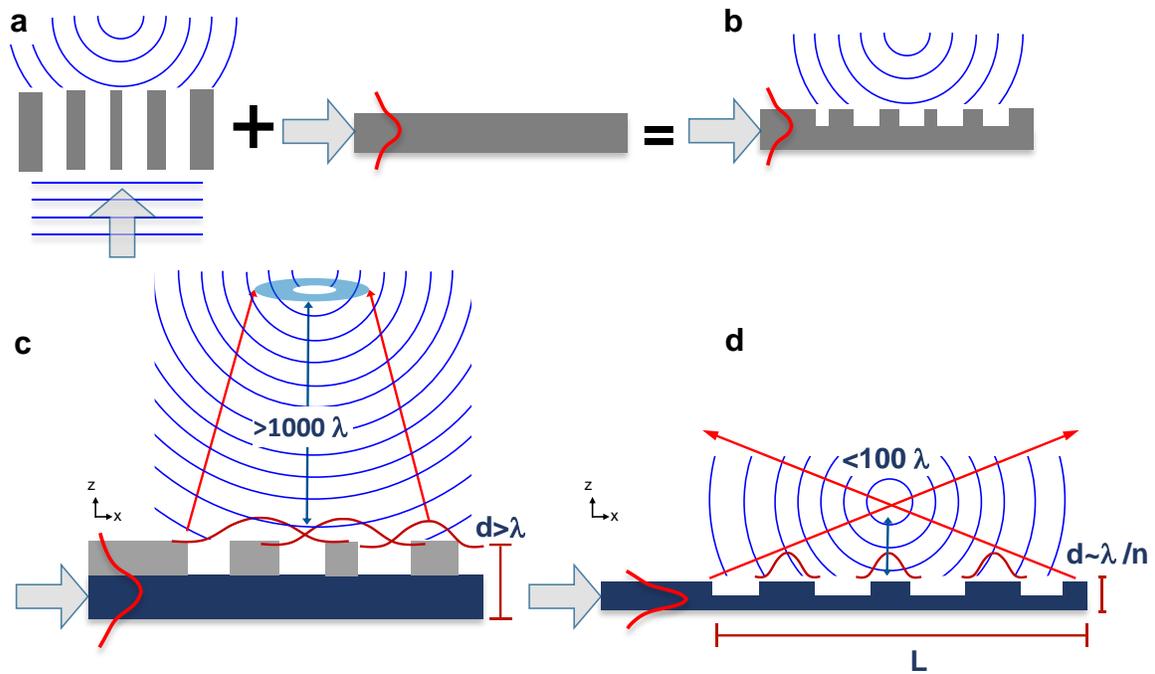

**Figure 1. Design of meta beam-shaper.**
**A)** Traditional free space metasurface that focus plane wave into a focal spot. **B)** Meta beam shaper integrated with a waveguide to feed the metasurface and arbitrarily shape the out-coupled beam. **C)** Current state-of-the-art low NA waveguide-fed metasurfaces targeting focal distances beyond 1000λ. Relatively large thickness, low confinement factor, and non-local fields are essential for inverse design using paraxial approximation. **D)** Design of high NA meta beam shaper operating at focal distances below 100λ requires high confinement single mode waveguide with local fields with highly non-paraxial beam conversion.

## RESULTS

**Design principles.**

Our goal is to design a waveguide-integrated meta beam shaper using a high index contrast dielectric to enable large deflection angles required for high NA applications. As opposed to non-local metasurfaces integrated into a low-confinement waveguide systems[9] (less than 5% confinement) we aim to implement waveguides with high confinement (over 90%) operating in a single mode regime. Single mode operation significantly eases the design of meta-atoms to be optimized to precisely shape just a single mode avoiding the complications of multi-mode optimization. High modal confinement and a high degree of locality can be beneficial[33,34] to provide large deflection angles and enable high NA applications. Additionally, design of meta beam shaper for a high index contrast high-confinement single-mode waveguide platform makes it compatible with versatile libraries of high-density integrated optical devices[35] for on-chip light manipulation, modulation, and detection.

To achieve a single mode operation (for a given polarization), the thickness of the guiding layer should be of the order of $\lambda/n$. That is significantly thinner than recent estimates of the minimal thickness $\lambda/(n-1)$ required for high NA free-space metasurfaces[33,36,37]. Long interaction distance of a traveling waveguide mode with a metasurface embedded in the waveguide core can potentially compensate for the limited thickness thus effectively overcome the fundamental trade-off[37] between achieving a desired high NA out-of-plane beams and the device thickness. Such single mode waveguide-fed meta beam shapers have been already demonstrated at thicknesses below the free-space limit[9]. Most of the design approaches are targeting long focal distances (1000$\lambda$) that justifies the use of paraxial approximation[9,12] and non-local distributed meta-atoms that significantly facilitates the inverse design (**Fig.1C**). However, paraxial approximation is no longer valid for the inverse design of meta beam shapers targeting highly convergent beams at focal distances shorter than 100$\lambda$ (**Fig.1D**). Instead of relying on non-local distributed resonances[9,12] in low-confinement waveguides, high NA applications require high-index contrast meta-atoms[31] that offer local control of high deflection angles[32] that make their inverse design challenging.

To overcome these challenges, we adopted a semi-intuitive forward design approach that consists of 3 consecutive steps (**Fig.2**). First, we design a standard metalens (**Fig.2A**) using a unit cell approximation for the target wavelength $\lambda$ and focal distance $f$ with meta-atoms arranged to produce a hyperbolic phase profile $2\pi/\lambda \left(f - \sqrt{f^2 + x^2 + y^2}\right)$ (Methods). To achieve full phase coverage and ensure that all rays arrive in phase at the focus, the metalens thickness is initially chosen as 6.2$\lambda/n$. Beam profiles calculated using 3D finite difference time domain (FDTD) for the metalens with NA of 0.37 illuminated with a plane wave (**Fig.2B**) demonstrate a tightly focused spot at the designed focal distance of 100 $\lambda$. The focusing quality is assessed by fitting a Gaussian profile to the central peak of the Airy disk taken along the corresponding axis. Full width at half maximum (FWHM) of the fitted Gaussians are 1.21 $\lambda$, 1.17 $\lambda$, and 8.89 $\lambda$, for the X-, Y- and Z-axis, correspondingly (black lines in **Fig.2G,H,K**). The corresponding volume of the ellipsoidal focal spot is 6.58$\lambda^3$. These numbers are close to a diffraction limited focusing approximated for the Fresnel near-field diffraction regime for a coherent light[38] (Methods) and are used as a reference for further comparisons.

Next, a standard single-mode (thickness ~ $\lambda/n$) waveguide grating is designed to outcouple and illuminate this free-space metalens (**Fig.2C**). This uniform grating is designed with a period, duty cycle, and etch depth to ensure a single channel diffraction into air and an intensity decay length along the x-axis (mode propagation direction) matching the metalens aperture (Methods).

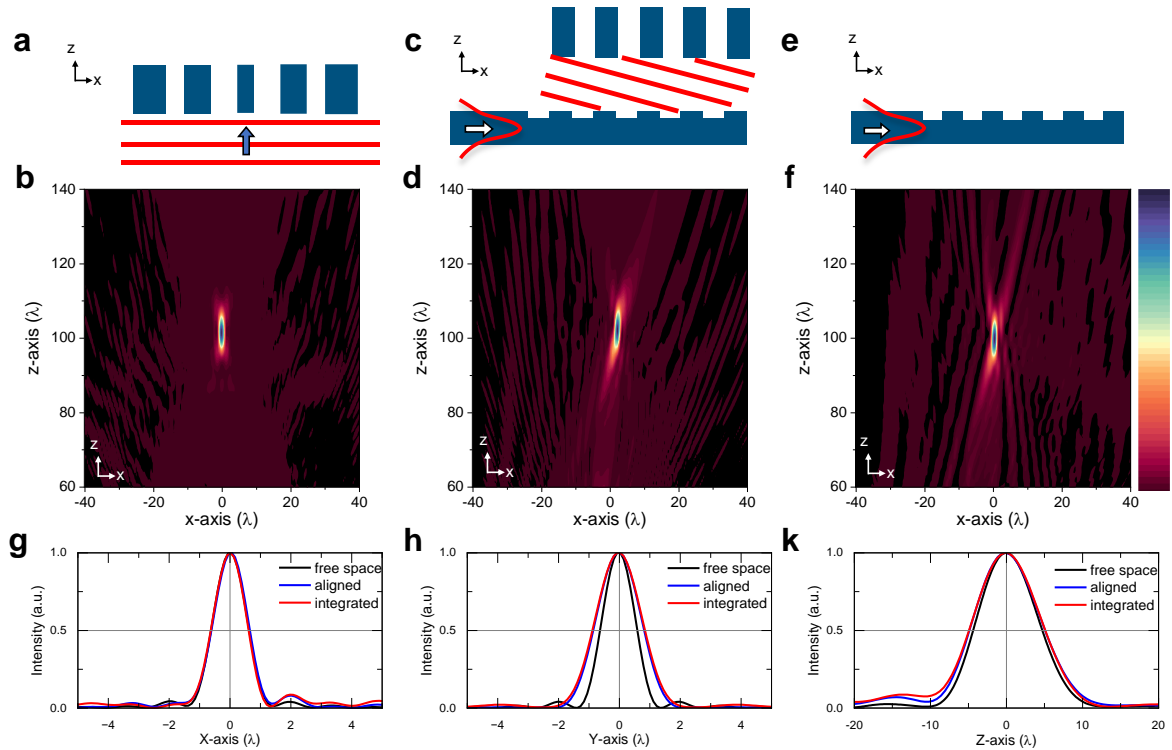

**Figure 2. Forward design of single mode meta beam shaper**
**A)** First design step considers standard free space metasurface fed by a planewave. **B)** Calculated light intensity above the device A with NA of 0.37 and focal distance of 100λ. Device plane is located at z=0μm. **C)** Second design step considers a metasurface placed on top of a single mode waveguide grating with meta-atoms shapes aligned with grating grooves. **D)** Calculated light intensity above the device C. **E)** Third design step considers an integrated meta beam shaper with meta-atoms partially etched into the core of a single mode waveguide. **F)** Calculated light intensity above the device E. **G)** Calculated normalized light intensity profile along x-axis for designs A (black), C (blue) and E (red). **H)** Calculated normalized light intensity profile along y-axis for designs A (black), C (blue) and E (red). **H)** Calculated normalized light intensity profile along z-axis for designs A (black), C (blue) and E (red).

These two components, a waveguide grating and a metalens, are placed on top of each other and the metalens library is re-optimized so that the meta-atoms of the metalens are aligned with grating teeth to produce a common vertical profile amenable for a single-step etching (**Fig.2C**). The metalens library is re-optimized to ensure that the phase of each meta-atom is modulated by altering its Y-width. To compensate for the grating diffraction angle $\theta_{gr}$ the metalens is re-patterned with a $2\pi/\lambda \left(f - \sqrt{f^2 + x^2 + y^2} - x \sin \theta_{gr}\right)$ phase profile. The total thickness of the integrated device is kept at 6.2λ/n. The resulting focal spot (**Fig.2D**) remains to be diffraction limited along the X- and Z-axis (blue lines in **Fig.2G,K**). The focus spot FWHM of 1.68 λ along the lateral Y-axis is 1.4 times wider than the diffraction limit presumably due to the non-uniformity of lateral illumination with 50% of a fundamental TE waveguide mode occupying about just half of the lateral metalens aperture. Such beam non-uniformity in the device plane results in broadening of corresponding point spread function (PSF) as they are related to each other via a Fourier transform[39].

Third, the device is thinned to a total thickness matching that of a single-mode waveguide λ/n (**Fig.2E**). The meta-atoms are patterned directly into the waveguide core to couple the waveguiding mode out of the plane and to impose phase shifts for each diffracting element to produce focusing. Maintaining high locality while illuminating all meta-elements as uniformly as possible determines the partial etch depth. As for device in **Fig.2C**, the etching depth and period of the integrated device in **Fig.2E** along the propagation direction (X-axis) are optimized for the duty cycle fixed at 50% until the decay constant of diffracted beam is equal to or greater than the metalens length while maintaining a single air diffraction channel. The width of periodically arranged meta-atoms along the Y-axis perpendicular to the propagation direction are altered to modulate the phase. This ultrathin configuration still forms a highly localized spot (**Fig.2F**) with X, Y, and Z axis profiles (red curves in **Fig.2G,K**) with FWHMs of 1.19λ , 1.68λ and 9.76λ, respectively, closely matching the response of a device in **Fig.2C**.

While not ideal, the integrated device produces a focal spot with a volume of 10.28λ³ at distances as small as 100λ. For light beams in the blue or ultraviolet wavelength range relevant to optogenetics[24] and to on-chip quantum optics with trapped ions[28-30] this would correspond to a volume below 1μm³ at focusing distances less than 50μm. These numerical experiments confirm our intuition that integrated meta beam shaper can yield tight focusing at a device thickness well below that of a free space metalens.

**Experimental design and fabrication**

Since we are targeting optogenetics applications[22-26], among others, the design of **Fig.2E** is optimized for 460nm wavelength. To achieve high confinement and single mode operation at this wavelength, a 250nm thick silicon nitride waveguide layer (n ≈ 1.89) is deposited on a 1μm thick silicon dioxide layer (n ≈ 1.44) on a silicon substrate (**Fig.3A**). To minimize absorption losses, the silicon nitride deposition using plasma-enhanced chemical vapor deposited (PECVD) is performed at 380°C using a mixed frequency recipe to achieve a nearly stoichiometric configuration[40] (Methods). A waveguide is formed by etching the SiNx layer down to silicon dioxide substrate with the waveguide width of 350nm to ensure a single TE mode with high 91% confinement factor and $n_{eff}$=1.77 (calculated mode profile in **Fig.3A**).

Meta atoms are defined by shallow 40nm etching into the waveguide core of 240nm (**Fig.3A**) that enables nearly uniform out-of-plane coupling along the propagation direction while maintaining high confinement of the fundamental mode. Short operating wavelength dictates the meta-atom sizes in the range between 60nm and 180nm (**Fig.3B**). Along the mode propagation direction (x-axis) meta-atoms are placed at a 270nm period with a 50% fill factor to satisfy the first-order Bragg condition for out-coupling. Perpendicular to the propagation direction (y-axis) meta-atoms are placed at 250nm period with their y-widths altered for phase modulation.

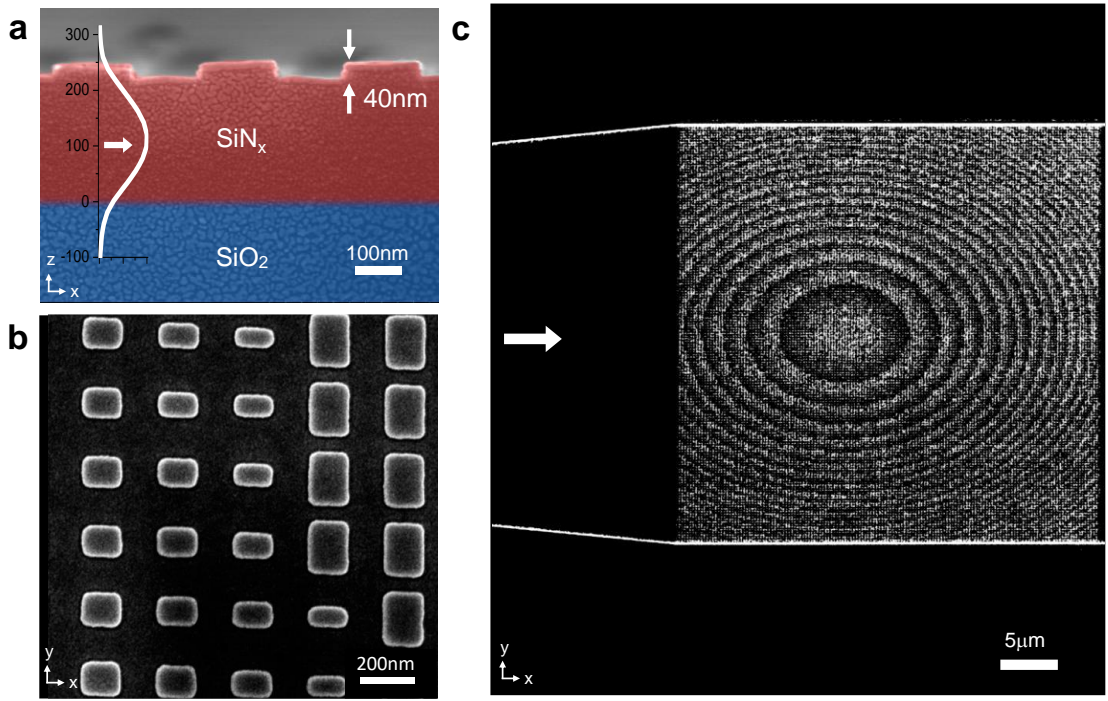

**Figure 3. Design and fabrication of single mode meta beam shaper.**
**A).** Cross-sectional SEM image of the fabricated waveguide-fed meta beam shaper. The meta-atoms are patterned by a 40nm partial etch into a 250nm $SiN_x$ film on top of a 1.1μm thick $SiO_2$ cladding. This thickness supports a single waveguide mode with over 90% confinement whose mode power density profile is overlaid. **B)** Top-down SEM image of meta-atoms patterned in the device x-y plane. Meta-atoms sizes are ranging from 60nm to 180nm to enable performance at wavelength of 473nm. **C)** Top-down SEM image of the meta beam shaper with NA 0.37 designed for focusing length of 50μm. This device area is 40μmx40μm fed with a linear taper partially seen on the left. Phase correction is applied to compensate for a slanted out-coupling angle and an axial astigmatism induced by a linear taper. Therefore, the zone center is shifted along the x-axis and phase zones are squeezed.

The design in **Fig.3C** corresponds to relatively small NA of 0.37 with targeted focal distance of 50μm above the chip surface that dictates the meta beam shaper size of 40μmx40μm. To feed the device, a fundamental TE mode is expanded adiabatically to 40μm device width using a linear taper (end of the taper is visible in **Fig.3C**). As the mode expands, however, it no longer stays collimated, and the phase fronts become curved inducing axial aberration (Supplementary information **SI Fig.S1**). To correct for these aberrations as well as for a slanted diffraction angle, a compensatory phase modulation is introduced into the metalens design (Methods). The metalens pattern in **Fig.3C** that includes both compensations is therefore shifted from the device center to ensure normal diffraction and the metalens phase zones are squeezed along the x-axis.

**Experimental results**

To measure diffracted beam profiles, the TE polarized light from a fiber-pigtailed 460nm laser diode is coupled to a single mode waveguide via a standard grating coupler and is fed into the meta beam shaper via a linear taper (Methods). To reconstruct the beam profile in a 3D, the images are acquired with a 100X objective with NA of 0.95 that is scanned in Z-axis perpendicular to the device plane at increments of 0.2μm starting from the device surface (z=0) to twice the focal distance above the chip (Methods).

**Fig.4A** shows beam intensity profile measured at the X-Z plane for a 40μmx40μm device of **Fig.3C** with designed NA of 0.37 and a focal distance of 50μm. Beams from different zones of the meta beam shaper converge to a tightly focused spot at a focal distance of 52μm and at the center of the device (x=0μm) close to the designed values. Beam profile in the Y-Z plane (**Fig.4B**) shows a symmetric tightly focused spot at the center of the device (y=0). Beam profile measured at the X-Y focal plane at z=52μm (**Fig.4C**) shows a nearly circular spot with characteristic Airy concentric ring ripples of much lower intensity. Analogous beam profiles calculated with 3D FDTD (Methods) are shown in **Fig.4E, F, G** for X-Z, Y-Z, and X-Y planes, respectively. Measured (black) and calculated (red) normalized intensity cross-sections through the focal spot along X-axis, Y-axis, and Z-axis are presented in **Fig.4D**. Gaussian fitting of experimentally measured cross-sections yields FWHM of 0.78μm, 1.02μm, and 5.41μm along the X, Y, and Z axes, respectively. Comparison with FDTD calculated FWHM for a device of **Fig.2F** indicates relatively good performance along the Y- and Z-axis and somewhat larger aberrations along the propagation X-axis. As a result, the experimental focal spot volume of 2.2μm$^3$ is about 2.2X larger than the calculated diffraction limited volume of 1μm$^3$.

To reach smaller focal volumes with larger NA of 0.7 dictates for the same focal distance of 50μm the device size to be increased to 100μmx100μm footprint with corresponding increase of the length of the feeding linear taper and recalculations of the phase compensation. The resulting measured beam profiles for the X-Z plane (**Fig.4H**), Y-Z plane (**Fig.4I**), and X-Y plane (**Fig.4J**) showed tightly focused spot at a focal distance of 48μm located at the device center as designed. Corresponding calculated profiles (**Fig.4L,M,N**) show similar structure of converging beams from different zones along X-Z plane (compare **Fig.4A** and **Fig.4H**) and Y-Z plane (compare **Fig.4B** and **Fig.4I**) as well as concentric Airy rings in the X-Y plane (compare **Fig.4C** and **Fig.4J**). Gaussian fitting of the beam cross-sections for both experimental (black) and calculated (red) curves along X, Y, and Z axes in **Fig.4K** yields FWHM of experimental spot size of 0.47μm, 0.45μm, and 1.63μm to be compared to the calculated FWHM of 0.31μm, 0.40μm, and 1.37μm for the same axes, respectively. The comparison confirms relatively good performance along Y- and Z- axis with just 10% broadening and somewhat larger aberrations along the X-axis similar to observed for NA of 0.37. Nevertheless, the resulting experimental focal spot volume of 0.18μm$^3$ is just 2X larger than the calculated volume of 0.09μm$^3$.

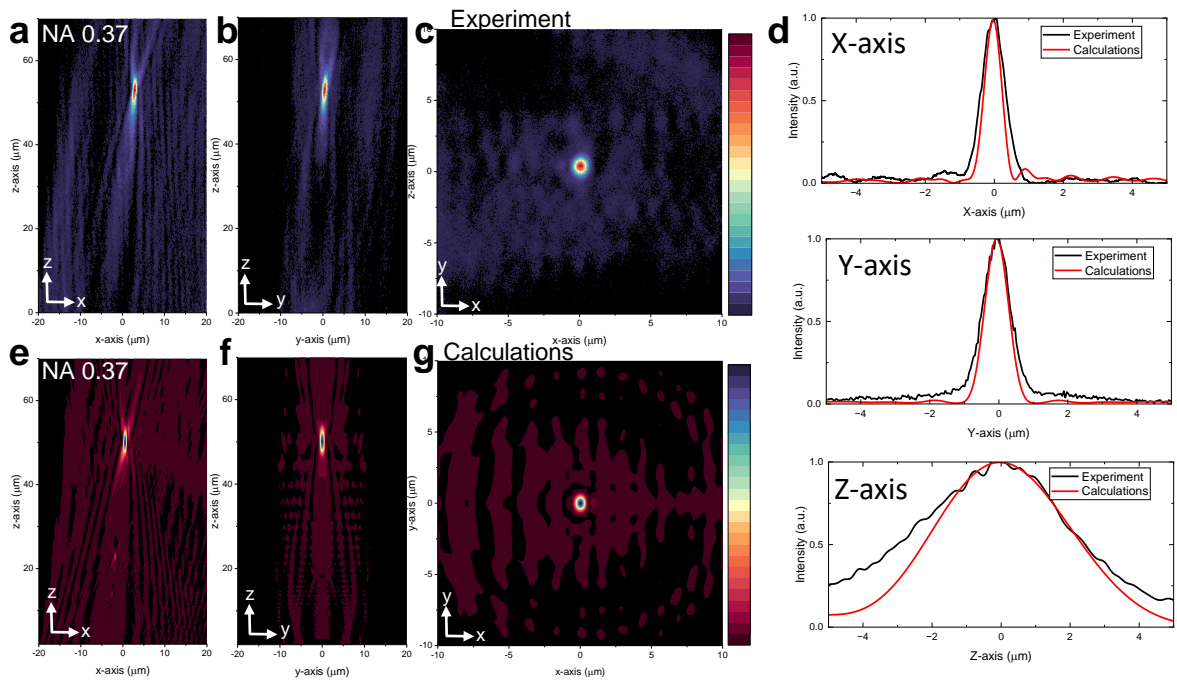
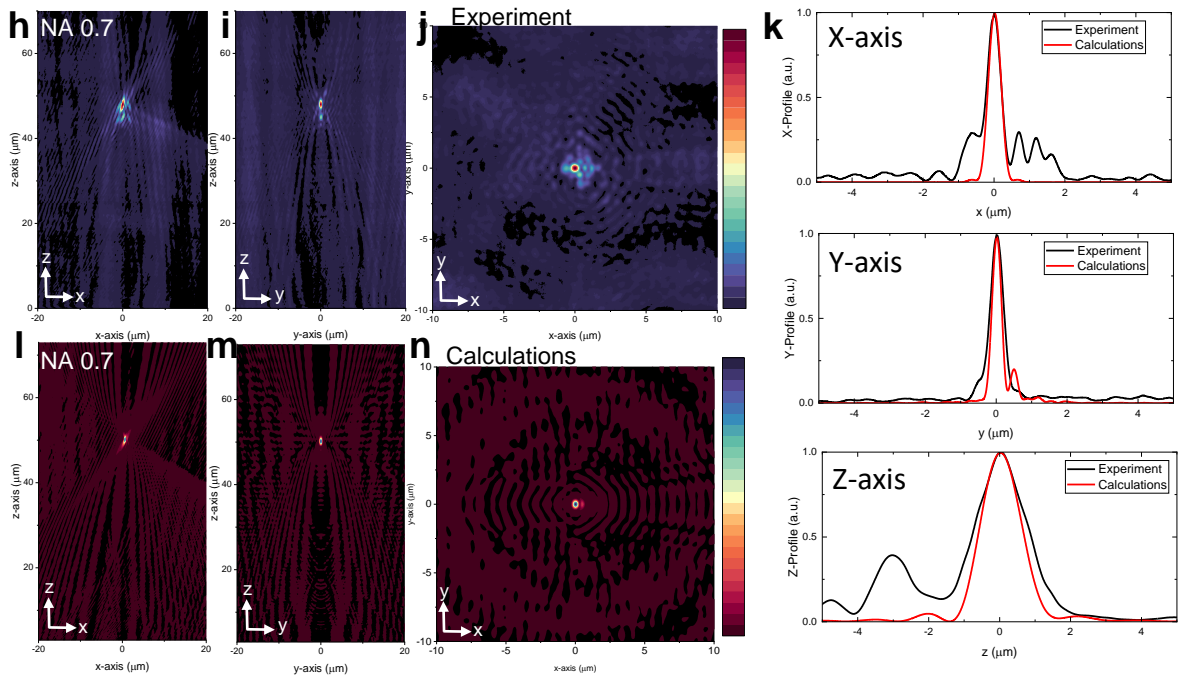

**Figure 4. Mode profiles for meta beam shaper with NA 0.37 and 50μm focal distance**
**A)** Measured intensity profile in the Z-X plane with z=0 corresponding to the device surface, and x=0 at the focal spot. Device size of 40 μm x40 μm with designed NA of 0.37 and a focal distance 50μm. **B)** Intensity profile in the Z-Y plane for the same device as in A). **C)** Intensity profile measured at the focal X-Y plane at z=52μm with 0,0 coordinates centered at the focal spot maximum for the same device as in A) and B). **D)** Normalized intensity cross-sections through the focal spot along X-axis, Y-axis, and Z-axis. Experimentally measured profiles (black curves) are compared to FDTD calculated profiles (red curves). **E)** FDTD calculated intensity profile in the Z-X plane. **F)** FDTD calculated intensity profile in the Z-Y plane. **G)** FDTD calculated intensity profile in the X-Y focal plane.

**H)** Measured intensity profile in the Z-X plane with z=0 corresponding to the device surface, and x=0 at the focal spot. Device size of 100 μm x100 μm with designed NA of 0.7 and a focal distance 50μm. **I)** Intensity profile in the z-y plane for the same device as in H). **J)** Intensity profile measured at the focal X-Y plane at z=48μm with 0,0 coordinates centered at the focal spot maximum for the same device as in H) and I). **K)** Normalized intensity cross-sections through the focal spot along X-axis, Y-axis, and Z-axis. Experimentally measured profiles (black curves) are compared to FDTD calculated profiles (red curves). **L)** FDTD calculated intensity profile in the Z-X plane. **M)** FDTD calculated intensity profile in the Z-Y plane. **N)** FDTD calculated intensity profile in the X-Y focal plane.

To approach focal volumes as small as $10^{-3} \mu m^3$ even higher NA above 0.95 is needed. To achieve such high NA while keeping the focal distance constant at 50μm, would require increase of the device size to a footprint above 300μm x 300μm. However, adiabatic tapering from a single mode waveguide with the width of just 350nm up to 300μm device size would require prohibitively large taper footprint. Instead, to achieve higher NA, we explored an alternative approach that maintains device footprint constant while reducing the focal distance. Following this approach, two separate series of devices were fabricated with footprints of 100μm x 100μm (**Fig.5A-G**) and 80μm x 80μm (**Fig.5H-N**).

For a series of devices with 100μm x 100μm footprint, the NA is scanned from 0.7 (**Fig.5A**) to as high as 0.95 (**Fig.5D**) with corresponding focal distance decreasing from 50μm to as small as 16μm. Corresponding mode profiles measured at the focal X-Y plane (**Fig.5A-D**) show tightly focused spots surrounded by characteristic concentric Airy rings. Corresponding focal spot cross-sections along the X- and Y-axis (bottom panel of **Fig.5A-D**) demonstrate, as expected, strong reduction of the FWHM of the central peak with decreasing the focal distance (increasing the NA) from 0.53μm x 0.38μm for 50μm focus distance (NA 0.7) to 0.36μm x 0.30μm for 16μm focus distance (NA=0.95). Similar tight focusing is observed for a series of devices with a footprint of 80μm x 80μm (**Fig.6E-H**) with the NA scanned from 0.37 to 0.95. The focal volume as small as 0.06μm$^3$ at a focal distance of just 13 μm is measured for a device with NA of 0.95 (**Fig.6H**).

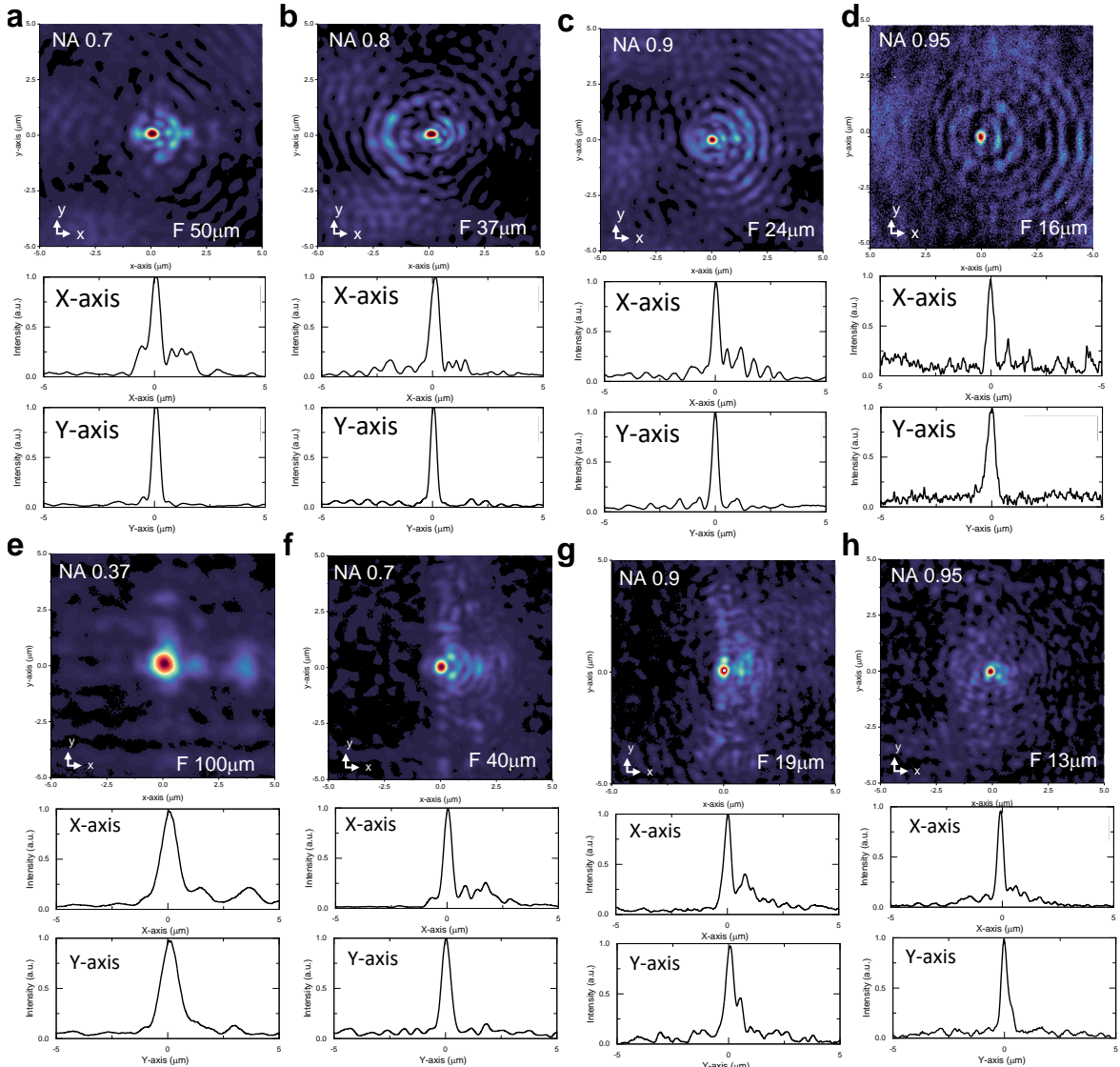

**Figure 5. Experimental mode profiles for devices with different footprint**
**A-D Top panel:** Measured intensity profiles in the x-y plane taken at corresponding focal distances for a series of devices with varying NA but constant 100μmx100μm. **A)** NA of 0.7, focal distance 51μm. **B)** NA of 0.8, focal distance 37μm. **C)** NA of 0.9, focal distance 24μm. **D)** NA of 0.95, focal distance 16μm. **Lower panel:** Corresponding cross-sections along X and Y axis taken from beam profiles on top.
**E-H Top panel:** Measured intensity profiles in the x-y plane taken at corresponding focal distances for a series of devices with varying NA but constant 80μmx80μm.
**E)** NA of 0.37, focal distance 100μm. **F)** NA of 0.7, focal distance 40μm. **G)** NA of 0.9, focal distance 19μm. **H)** NA of 0.95, focal distance 13μm. **Lower panel:** Corresponding cross-sections along X and Y axis taken from beam profiles on top.

**DISCUSSION AND OUTLOOK**

The measured FWHMs of focal spots along all three axes are summarized in **Fig.6A,B,C** for all the devices. They are compared to the FWHM calculated with the 3D FDTD (dashed red curves) and with the FWHM for the diffraction limited focused spot (dashed black curves) calculated for a Fresnel near-field diffraction regime (Methods). For all the devices, the measured the focal spot size is decreasing drastically with the increase of the NA as expected from the design. Even in the presence of noticeable aberrations the focal volume of the spot (**Fig.6D**) is decreasing from 2.23μm$^3$ for NA of 0.37 at a focal distance of 50μm to as small as just 0.06μm$^3$ at a focal distance of 13μm measured for a device with NA of 0.95.

The design approach we are taking enables exploration of even smaller focal volumes approaching 10$^{-3}$ μm$^3$ by tuning the NA and device footprints. **Fig.6E** shows estimated diffraction limited volumes that can be achieved for a given focal distance by changing the device footprint. For example, focal volumes as small as 2·10$^{-3}$μm$^3$ at focal distance of 20μm (black line in **Fig.6E**) can be achieved with a device aperture of 150μm that can be illuminated with an adiabatic taper of a reasonable size. It would be difficult to achieve such strong focusing for larger focal distances of 50μm and 100μm (red and blue lines in **Fig.6E**) as the device aperture will exceed reasonable 300μm (red dotted line in **Fig.6E**). Decreasing the device footprint below 30μm x 30μm (dotted red line in **Fig.6E**) will ease tapered coupling, however with the potential downside of increasing contribution of edge diffraction and diminishing number of meta-atoms (below 10$^4$) available for shaping the beam that eventually will broaden the PSF.

To achieve these ultra-small volumes, axial and lateral aberrations in the current design need to be addressed. Most significant aberrations are observed for the experimentally measured spot size (circles) along the X-axis (**Fig.6A**) that is about 1.81 ± 0.15 (mean ± SEM) times larger than expected from the FDTD calculations (red dashed curve) and the diffraction limit (black dashed curve) for all the NA and device footprints. Such broadening of the PSF can be related to non-uniform filling of the aperture due to scattering losses that the waveguide mode encounters while propagating along the X-axis that worsens already appreciable beam non-uniformity. Random scattering of in-plane mode not only affects its amplitude, but also inevitably generates phase errors that accumulate during the mode propagation. Indeed, while over 10dB suppression of the Airy ring patterns in comparison with the central peak is prominent for the Y-axis (**Fig.5A-D,** lower panel), the Airy rings along the X-axis have higher amplitudes with just 5dB difference with respect to the central peak especially visible at positive X-coordinates after the mode passes the central zone (**Fig.5A-D,** lower panel). These observations, observed also for the devices with 80μm x 80μm footprint (**Fig.5E-H**), indicate an incomplete interference forming the Airy ring pattern that is affected by accumulated phase errors.

Experimentally measured FWHM for Y-axis (circles in **Fig.6B**) are within just one standard deviation (SD of ±0.15) from the FDTD calculations (red dashed line). Even tighter lateral focusing close to the diffraction limit (black dashed line) can potentially be achieved by redesigning the adiabatic taper and meta-atoms library to minimize the effect of non-uniform filling of the device lateral aperture. With relatively good axial focusing along Z-axis (**Fig.6C**) that is close to the FDTD calculations and the diffraction limit, the focal volume (**Fig.6D**) is mostly affected by the X-axis and can be addressed with better control of fabrication tolerances to minimize scattering loss and accumulation of phase errors during mode propagation in the device plane.

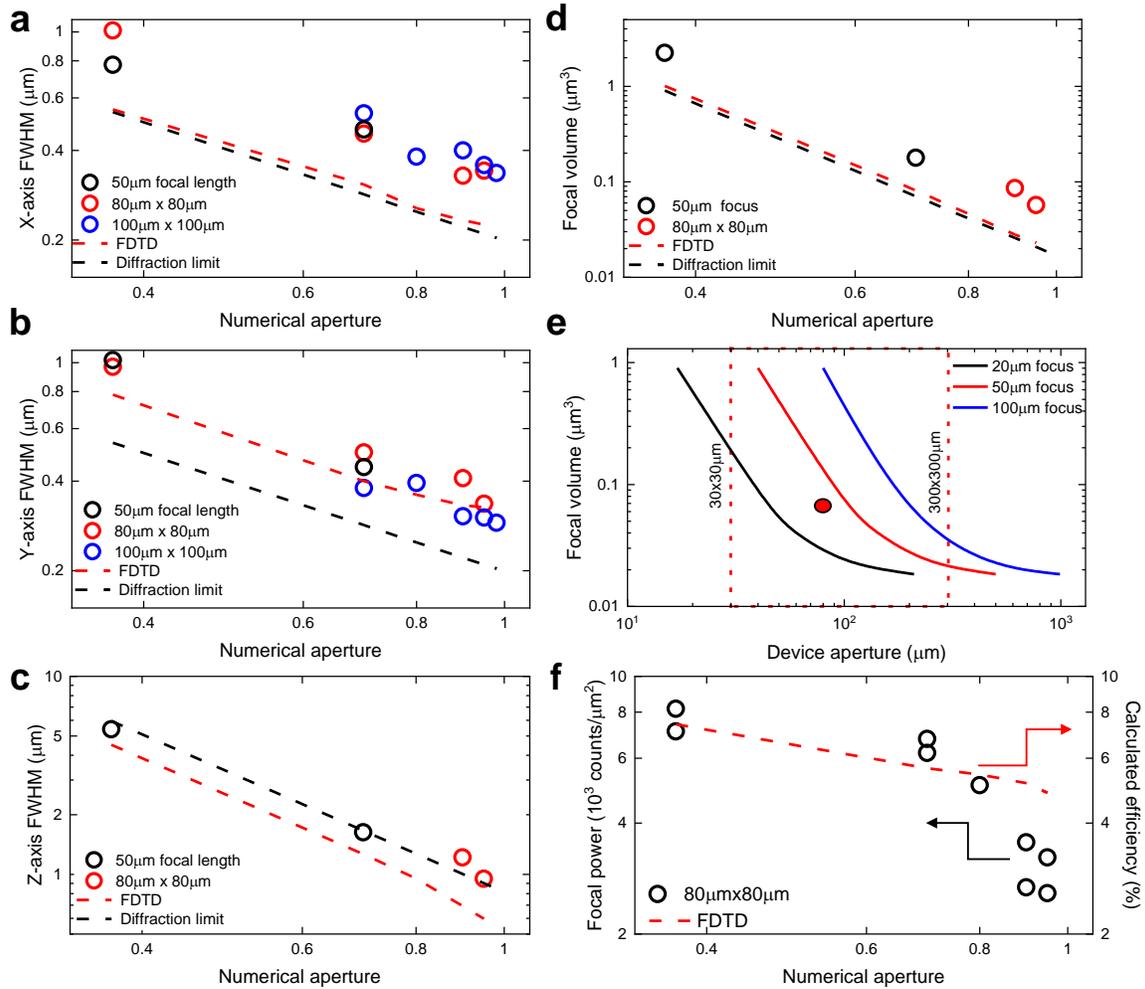

**Figure 6. Focusing quality and efficiency**
**A)** X-axis FWHM of a focused spot for a series of devices with different NA. Black dots correspond to devices from Fig.4 with constant focal distance of 50μm but varying footprint. Red circles correspond to 80μmx80μm footprint from Fig.5E-H. Blue circles correspond to 100μmx100μm footprint with varying NA from Fig.5A-D. Dashed red line corresponds to FDTD calculated values. Dashed black line corresponds to diffraction limit. **B)** Y-axis FWHM for the same devices as in A). Dashed red line corresponds to FDTD calculated values. Dashed black line corresponds to diffraction limit. **C)** Z-axis FWHM for the same devices as in A). Dashed red line corresponds to FDTD calculated values. Dashed black line corresponds to diffraction limit. **D)** Focused spot volume from FWHMs for measured (circles) and calculated with FDTD (red dashed line) and using diffraction limits (black dashed line). **E)** Log-log plot of the estimated diffraction limited focused spot volume as a function of a device size. Black, red, and blue curves correspond to fixed focal distances of 20μm, 50μm, and 100μm. Red circle indicate experimentally measured volume for a device in Fig.5H. **F)** Measured optical power (left vertical axis) within 3μmx3μm region taken around the focal spot at the focal x-y plane for a series of devices with 80μmx80μm footprint. Dashed red line corresponds to the right vertical axis is FDTD calculated focusing efficiency (power at the focal spot divided by the total power within a focal plane).

The calculated focusing efficiency of the designed meta beam shaper is about 8% for a small NA of 0.37 (right vertical axis in **Fig.6F**) with slow decrease to higher NA. However, even for NA as large as 0.95 the calculated focusing efficiency is still appreciable 5% limited primarily by a small extraction efficiency of the grating that is designed to provide a nearly uniform mode amplitude along the device length. Experimentally, power delivered by the device to the focused spot is measured for a series of devices with 80μm x 80μm footprint (**Fig.6F)** by integrating the light intensity within a 3μm x 3μm area around the focal spot at the focal X-Y plane. With the increase of the NA, the power is slowly decreasing following the predicted trend until a significant 2X drop for NA higher than 0.8 most likely due to experimental limitations (the NA of the objective lens, resolution of the camera, focus scanning step). Larger efficiency is expected with the increase of the partial etch step that would require adiabatic apodization to maintain lateral beam uniformity to avoid X-axis PSF broadening.

Demonstrated high-NA integrated meta beam shapers with focal volumes below 1μm$^3$ at the ultra-short focal distances from the surface of the integrated chip can potentially be used for atom ionization, trapping, cooling, quantum state manipulation, and state readout from many ions in parallel[28,29,35]. For example, quantum state measurement of a trapped Sr$^+$ ions is typically accomplished by collecting resonant fluorescence at 422nm using high-NA bulk optics[41]. Integrated high-NA meta beam shaper can be used instead to collect emitted photons directly at the individual ion trap site and couple them to a waveguide integrated photodetector for fast readout making a scalable parallel architecture feasible. High-NA optical tweezers based on our meta beam shapers can focus light into ultra-small volumes near the chip surface that might enable improved cold-atom transport[30] and high-fidelity entanglement[42].

Beyond quantum optics high-NA integrated meta beam shapers can also find applications in high resolution optical imaging inside the biological tissues[26] and optical manipulation of neural activity via optogenetic stimulation[22,23,25]. For example, typical optogenetic stimulation often result in a wide spread of excitation or inhibition beyond the targeted neurons due to strong coupling between neurons[24]. However, if the excited volume is highly localized via high NA meta beam shaper, a single targeted neuron can be individually addressed due to exponential dependence of the evoked spike rate on delivered optical power[24].

# METHODS

**Grating design.** To feed the metalens a grating is designed in dielectric stack composed of a 250nm thick silicon nitride waveguiding layer, silicon dioxide bottom cladding and air top cladding as shown in **Fig. 2C,D**. The grating is designed to have a single channel of diffraction into the air and an intensity decay within 1/e of the initial intensity along the length of the grating (x-axis). The former is required to ensure a single diffraction order and the latter is needed to fill the aperture of the metalens as uniformly as possible. 3D FDTD calculations of the grating are performed in Lumerical FDTD (*Ansys, Inc*).

The source is the fundamental TE mode calculated for a given grating width and silicon nitride film thickness which is launched into the structure. This mode propagates for four wavelengths before encountering the grating region. A field monitor placed 2 wavelengths above the waveguide plane and spanning the entire device area captures the scattered electric and magnetic fields. The FDTD volume encloses the waveguiding layer completely and has perfectly absorbing boundary conditions. The intensity decay in the grating is captured by the field monitor and is fitted to an exponential decay to obtain the decay constant. The partial etch depth and period of the grating are swept keeping the duty cycle fixed at 50% until the decay constant is equal to or greater than the metalens length while maintaining a single air diffraction channel. This optimization suggests a period of 270nm and a 40nm partial etch depth for a 250nm thick waveguiding film.

**Metalens design.** Detailed design strategy is discussed in the text. Initially, a standard metalens library is computed in a unit cell approximation for the target wavelength λ with meta-atoms arranged to produce a hyperbolic phase profile $\phi(x,y) = 2\pi/\lambda \left(f - \sqrt{f^2 + x^2 + y^2}\right)$ for diffraction-limited focusing. To achieve full phase coverage and ensure that all rays arrive in phase at the focus, the metalens thickness is initially chosen as 6.2λ/n. In the next step, the calculations are repeated for a metalens being fed with a grating input profile instead of a plane wave. Next, the meta-atoms of a metalens are aligned to the grating teeth and these two components, a waveguide grating and a metalens, are placed on top of each other. The metalens library is re-optimized to ensure that the phase of each meta-atom is modulated by altering its y-width. Finally, the total thickness of the meta-atoms is reduced to that of a single mode waveguide (250nm) and meta-atoms are defined by shallow etching into the waveguide core for a depth defined by the grating design (40nm). After the simulation is completed, the Lumerical function 'farfieldexact3d' is used to project the collected monitor data into the far field pattern. This projection spans the area of the device and goes to twice the focal length in height. The field data is normalized to the maximum and orthogonal cross-sections in the x-z, y-z, and x-y planes of this 3D volume are then used to characterize the focused spot.

**Tilt Correction.** In an integrated meta beam shaper, the output beam might not necessarily be normal to the device plane. To ensure that the spot formed by the metalens is still centered above the device, we compensate for this tilt by adding an additional phase term to the meta-atoms design which compensates for the phase shift along the tilt direction. The amended phase profile is:

$$\phi(x,y) = \frac{2\pi}{\lambda}\left(f - \sqrt{f^2 + x^2 + y^2} - x \sin\theta_{tilt}\right) \qquad \text{Eq.1}$$

**Taper-induced phase front aberration correction.** When the fundamental mode is tapered out from a single mode waveguide (350 nm width) to the footprint of the meta beam shaper (width as large as hundreds of microns) the phase front is distorted as shown in **SI Fig.S1**. To estimate the wavefront curvature, we assume the single-mode waveguide to be of negligible width compared to the metasurface (point source approximation), hence along the linear taper, the single-mode will expand into circular wavefronts. For a linear taper with a length *L*, the phase difference the propagating mode will acquire at the entrance to the meta beam shaper between its center and a meta atom located at y-distance along its width will be $\Delta\phi = \frac{2\pi n_{eff}}{\lambda}\left(\sqrt{L^2 + y^2} - L\right)$ where $n_{eff}$ is the effective index of the fundamental waveguide mode. The same effect will occur at both the input and output grating couplers. The taper length and mode index are adjusted accordingly. As the input grating coupler is much smaller than the output grating coupler, we assume a one-to-one correspondence between the points along the input grating coupler width and points along the metasurface width. Therefore, for every y along the metasurface width, $y_{input} = \frac{\text{Input grating width}}{\text{Metasurface width}} \times y$. The corresponding phase profile with the grating angle and taper correction is:

$$\Delta\phi(x,y) = \frac{2\pi}{\lambda}\left(f - \sqrt{f^2 + x^2 + y^2} - x\sin\theta_{grating} - n_{input}\left(\sqrt{L_{input}^2 + y_{input}^2} - L_{input}\right) - n_{output}\left(\sqrt{L_{output}^2 + y^2} - L_{output}\right)\right) \qquad \text{Eq.2}$$

The output in the above equation is the integrated metasurface, and the input is the input grating coupler.

**Fabrication.** The waveguiding layer is a 250nm thick layer of plasma-enhanced chemical vapor deposited (PECVD) silicon nitride (*Oxford Instruments, Inc.*) deposited at 380°C using a mixed frequency recipe with a NH3:SiH4 gas ratio of 2.5:1 to achieve a nearly stoichiometric configuration needed to minimize losses in the visible[8]. This film is deposited on a bottom cladding of PECVD $SiO_2$ supported on a Si substrate.

Refractive indices for silicon nitride and silicon dioxide are based on that obtained for 3-wavelength ellipsometry data. The data points are fit to the Cauchy refractive index formula to obtain corresponding Cauchy coefficients. These are used to calculate the refractive index for the wavelength of interest and are used as parameters for the simulation.

The first lithography step is to define the single-mode waveguide, the slab waveguide region for the metasurface and input grating coupler and the tapers connecting them together. These are patterned on the silicon nitride by 150KeV electron beam (e-beam) patterning system (*Elionix ELS-G150*) of a 470nm thick layer of CSAR e-beam photoresist, followed by a timed plasma etch (*Oxford Instruments, Inc.*) to clear the silicon nitride thickness and expose the oxide underneath. A trench width of 3mm is used to isolate the waveguide from the remaining film as CSAR is a positive tone e-beam resist. A second layer of aligned e-beam lithography with a 200nm thick CSAR resist followed by a timed 40nm etch defines the meta-elements within the waveguide. Suitable shape-based proximity effect correction measures (*BEAMER, GenISys GmbH*) are taken to minimize the deviation between the designed and fabricated meta-element dimensions. After each etch step, the resist is removed using a resist stripper (*Microposit 1165*) at 80°C and cleaned further by O2 plasma ashing.

**Experimental Setup.** The laser source is a 460nm diode laser (*Wave Spectrum*) pigtailed to a single mode fiber. After passing through a fiber polarization controller laser light is coupled into a chip via a standard grating coupler with the output tapered to a width of a single mode waveguide and is scattered out and shaped by the integrated metalens. A 100x NA 0.9 objective (*Olympus MS-Plan F180*) attached to a 24 Megapixel DSLR camera (*Canon Rebel T7*) with their optical axis perpendicular to the chip surface to characterize the scattered beam quality. To reconstruct the beam shape in 3D, the objective and camera are mounted on a motorized stage. Images are taken at increments of 1μm starting from the device surface to twice the focal length above the chip.

**Data Processing and analysis.** A custom pipeline written in Python takes the set of images collected by the experimental setup, spatially calibrates the image set in all three dimensions and creates a 3D reconstruction of the beam. Three orthogonal cross-sections of the spot in the x-z, y-z, and x-y planes are extracted. A Gaussian is fit to the central peak of the Airy intensity profile along each direction. The full width at half maximum (FWHM) of the fitted Gaussian is taken to be the measure of the spot size.

**Estimation of diffraction limits.** In the experiment, measurements of the beam waist (at $1/e^2$ amplitude) or the Airy disk diameter (first nulls of the Airy function) are less reliable than measurements of the full width at half maximum (FWHM). Therefore, the common estimates of diffraction limits that typically are derived for the beam waist or the diameter of the Airy disk need to be recalculated to the FWHM to be compared with experimental data. Our devices are designed for the high NA and short focal distances that correspond to a Fresnel near-field diffraction regime where Fresnel number $A^2/z\lambda$ ($A$ is the metalens width and $z$ is the focal distance) is larger than 1. For this "near-field imaging regime" the lateral resolution limit[38] (width corresponding to the first nulls of the Airy function) for a coherent light can be estimated as $\lambda/NA$. Recalculating to the FWHM of the Gaussian curve fitted to the Airy function produces lateral diffraction limited spot size with FWHM of

$$\text{FWHM}_{\text{lat}} = 0.436 \cdot \lambda/NA \qquad \text{Eq.3}$$

For the axial diffraction limit the Rayleigh range formulated for a coherent light at the "near-field imaging regime"[38] implies that the first zero of the axial intensity distribution function must occur at $z = \frac{2\lambda}{NA^2}$. The axial light intensity along the focal Z-axis could not be described by an Airy function but rather by an equation[43] $I(z) = I_0 \left(\frac{\sin u/4}{u/4}\right)^2$ where $u = \frac{2\pi}{\lambda}(NA)^2 z$. This function has a first zero when $u/4 = \pi$. Its amplitude falls by half when $u/4 = \pm 1.3916$. The corresponding half maximum point will then be $z = \frac{1.3916}{\pi}\frac{2\lambda}{NA^2}$. Therefore, the diffraction limited FWHM is:

$$\text{FWHM}_{\text{axial}} = 2 \cdot \frac{1.3916}{\pi}\frac{2\lambda}{NA^2} \qquad \text{Eq.4}$$

The lateral and axial diffraction limits estimated using these equations Eq.3 and Eq.4 are presented in **Fig.6 A-D** (black dashed lines). The focal volume is estimated as a volume of corresponding ellipsoid:

$$V = \frac{4}{3}\pi \left(\frac{\text{FWHM}_x}{2}\right)\left(\frac{\text{FWHM}_y}{2}\right)\left(\frac{\text{FWHM}_z}{2}\right) \qquad \text{Eq.5}$$

**ACKNOWLEDGEMENTS**

Research reported in this publication was supported in part by the NINDS of the NIH grants UF1NS107677 and RF1NS126061.

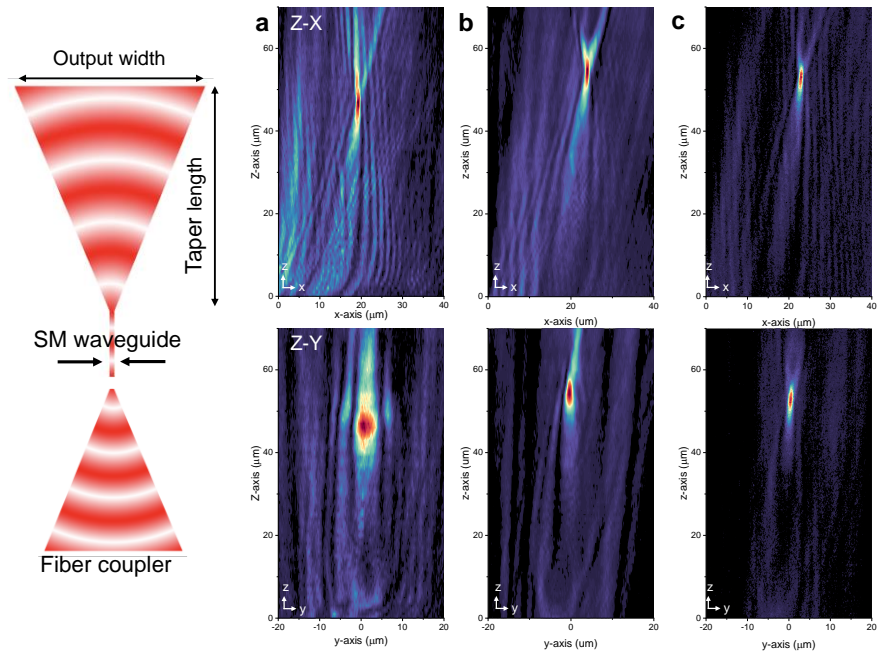

**Figure S1. Compensation of taper-induced phase distortion**
**Left:** Schematics of the phase front distortion of the waveguide mode propagating from a fiber coupler to the single mode waveguide that is coupled to a linear adiabatic taper to match the final device width. The phase front exhibits axial distortion with curvature dependent on the taper width and length.
**A)** Beam profiles in X-Z plane (upper panel) and Y-Z plane (lower panel) measured for a meta beam shaper with NA 0.37, focal distance 50μm, and a footprint of 40μm x 40μm. The device is being fed with a taper with a length of 200μm. Meta beam shaper design does not include the phase front correction. **B)** Analogous measurements for analogous device with phase front corrections implemented. **C)** Beam profiles for a device fed with 1000μm long taper and with phase corrections implemented.